\pdfoutput=1
\documentclass[%
 reprint,
 superscriptaddress,
 amsmath,amssymb,
 floatfix,
]{revtex4-2}

\usepackage{sidecap}
\usepackage{hyperref}
\usepackage{xcolor}
\usepackage{siunitx}
\usepackage{graphicx}

\begin{document}                  

\title{Three-Dimensional Coherent Diffractive Imaging of Isolated Faceted Nanostructures}

\author{Alessandro Colombo}
\email{alcolombo@phys.ethz.ch}
\affiliation{Laboratory for Solid State Physics, ETH Zurich, 8093 Zurich, Switzerland}

\author{Simon Dold}
\affiliation{European XFEL GmbH, 22869 Schenefeld, Germany}

\author{Patrice Kolb}
\affiliation{Laboratory for Solid State Physics, ETH Zurich, 8093 Zurich, Switzerland}

\author{Nils Bernhardt}
\affiliation{Technische Universität Berlin, Institut für Optik und Atomare Physik, 10623 Berlin, Germany}

\author{Patrick Behrens}
\affiliation{Technische Universität Berlin, Institut für Optik und Atomare Physik, 10623 Berlin, Germany}

\author{Jonathan Correa}
\affiliation{Deutsches Elektronen-Synchrotron DESY, 22607 Hamburg, Germany}

\author{Stefan D{\"u}sterer}
\affiliation{Deutsches Elektronen-Synchrotron DESY, 22607 Hamburg, Germany}

\author{Benjamin Erk}
\affiliation{Deutsches Elektronen-Synchrotron DESY, 22607 Hamburg, Germany}

\author{Linos Hecht}
\affiliation{Laboratory for Solid State Physics, ETH Zurich, 8093 Zurich, Switzerland}

\author{Andrea Heilrath}
\affiliation{Technische Universität Berlin, Institut für Optik und Atomare Physik, 10623 Berlin, Germany}

\author{Robert Irsig}
\affiliation{Department of Physics, University Rostock, 18051 Rostock, Germany}

\author{Norman Iwe}
\affiliation{Department of Physics, University Rostock, 18051 Rostock, Germany}

\author{Jakob Jordan}
\affiliation{Technische Universität Berlin, Institut für Optik und Atomare Physik, 10623 Berlin, Germany}

\author{Björn Kruse}
\affiliation{Department of Physics, University Rostock, 18051 Rostock, Germany}

\author{Bruno Langbehn}
\affiliation{Technische Universität Berlin, Institut für Optik und Atomare Physik, 10623 Berlin, Germany}

\author{Bastian Manschwetus}
\affiliation{Deutsches Elektronen-Synchrotron DESY, 22607 Hamburg, Germany}

\author{Franklin Martinez}
\affiliation{Department of Physics, University Rostock, 18051 Rostock, Germany}

\author{Karl-Heinz Meiwes-Broer}
\affiliation{Department of Physics, University Rostock, 18051 Rostock, Germany}

\author{Kevin Oldenburg}
\affiliation{Department of Physics, University Rostock, 18051 Rostock, Germany}

\author{Christopher Passow}
\affiliation{Deutsches Elektronen-Synchrotron DESY, 22607 Hamburg, Germany}

\author{Christian Peltz}
\affiliation{Department of Physics, University Rostock, 18051 Rostock, Germany}

\author{Mario Sauppe}
\affiliation{Laboratory for Solid State Physics, ETH Zurich, 8093 Zurich, Switzerland}

\author{Fabian Seel}
\affiliation{Technische Universität Berlin, Institut für Optik und Atomare Physik, 10623 Berlin, Germany}

\author{Rico Mayro P. Tanyag}
\affiliation{Technische Universität Berlin, Institut für Optik und Atomare Physik, 10623 Berlin, Germany}

\author{Rolf Treusch}
\affiliation{Deutsches Elektronen-Synchrotron DESY, 22607 Hamburg, Germany}

\author{Anatoli Ulmer}
\affiliation{Technische Universität Berlin, Institut für Optik und Atomare Physik, 10623 Berlin, Germany}

\author{Saida Walz}
\affiliation{Technische Universität Berlin, Institut für Optik und Atomare Physik, 10623 Berlin, Germany}

\author{Thomas Fennel}
\affiliation{Department of Physics, University Rostock, 18051 Rostock, Germany}

\author{Ingo Barke}
\affiliation{Department of Physics, University Rostock, 18051 Rostock, Germany}

\author{Thomas M{\"o}ller}
\affiliation{Technische Universität Berlin, Institut für Optik und Atomare Physik, 10623 Berlin, Germany}

\author{Bernd von Issendorff}
\affiliation{Department of Physics, University of Freiburg, 79104 Freiburg, Germany}

\author{Daniela Rupp}
\affiliation{Laboratory for Solid State Physics, ETH Zurich, 8093 Zurich, Switzerland}

\begin{abstract}
The structure and dynamics of isolated nanosamples in free flight can be directly visualized via single-shot coherent diffractive imaging using the intense and short pulses of X-ray free-electron lasers. Wide-angle scattering images even encode three-dimensional morphological information of the samples, but the retrieval of this information remains a challenge. Up to now, effective three-dimensional morphology reconstructions from single shots were only achieved via fitting with highly constrained models, requiring a priori knowledge about possible geometrical shapes. Here we present a much more generic imaging approach. Relying on a model that allows for any sample morphology described by a convex polyhedron, we reconstruct wide-angle diffraction patterns from individual silver nanoparticles. In addition to known structural motives with high symmetries, we retrieve imperfect shapes and agglomerates which were not accessible previously. Our results open new routes towards true 3D structure determination of single nanoparticles and, ultimately, 3D movies of ultrafast nanoscale dynamics.
\end{abstract}

\maketitle

\section{Introduction}
Coherent Diffractive Imaging (CDI) is a lensless technique that exploits the interference effects of coherent radiation scattered by an isolated sample to retrieve its spatial properties \cite{miao1999extending, chapman2010coherent, miao2011coherent, seibert2011single}. A highly intense light beam intercepts the specimen of interest, and the intensity of the diffracted light is collected by a 2D-detector in \emph{far-field} condition. CDI does not make use of optical devices, and its achievable spatial resolution is, in principle, only limited by the radiation wavelength. Thus, it is capable of fully exploiting the high-intensity ultra-short light pulses provided by the recently available Free-Electron Lasers (FELs) to image isolated nanosamples.
Examples range from biologically relevant specimen such as single viruses \cite{seibert2011single,ekeberg2015three} or macromolecules \cite{ekeberg2022observation} to atomic clusters \cite{rupp2012identification}, nanocrystals \cite{barke20153d}, and even such fragile and short lived structures as aerosols \cite{bogan2010aerosol} or superfluid helium nanodroplets \cite{langbehn2018three}.
The extremely short pulses further allow time-resolved imaging of ultrafast dynamics in nanoscale structures which has opened up unprecedented possibilities for light-matter interaction studies \cite{bostedt2012ultrafast,fluckiger2016time, Rupp2020, peltz2022few}.

CDI experiments can be roughly divided into two regimes defined by the angle upon which diffraction signal can be acquired. In the \emph{small-angle} regime, the maximum scattering angle reaches up to only a few degrees, and correspondingly the maximum transferred momentum is much smaller than the wave vector of the incident radiation \cite{barke20153d}. In this case, the diffraction pattern is proportional to the square amplitude of the Fourier Transform (FT) of the two-dimensional sample density, projected onto the plane orthogonal to the beam \cite{guinier1955small}. The field phase, lost in the acquisition process, can be effectively recovered by so-called \emph{phase retrieval} algorithms, which then allow the reconstruction of the sample’s projected density \cite{sayre1952some, fienup1987phase, marchesini2003x, loh2012fractal,seibert2011single,pedersoli2013mesoscale}. Three-dimensional reconstruction of samples can also be achieved in the small-angle range by tomographic approaches, i.e., by combining 2D projections of the same object or identical replicas in different orientations \cite{miao2006three, jiang2010quantitative, lundholm2018considerations, loh2010cryptotomography}. However, for the investigation of isolated systems with intense FEL pulses, this approach is practicable only in special cases \cite{loh2009reconstruction, ekeberg2015three, xu2014single} and unsuitable for dynamic investigations.

Three-dimensional features of the sample are, on the other hand, naturally encoded in a single diffraction image acquired in the \emph{wide-angle} regime. In this case, the maximum acquired transferred momentum is comparable in magnitude to the wavevector of the incoming and scattered photons. In particular, the component along the beam propagation direction is no longer negligible and, consequently, also partial 3D morphological information is imprinted in a single diffraction shot \cite{barke20153d}. This advantageous feature is, however, counterbalanced by some drawbacks. First of all, the scattering cross section of the sample material and/or the intensity of the radiation have to be sufficiently high to provide useful scattering signal at high scattering angles. When such an experimental condition is met, the extraction of the 3D morphological information from the single diffraction pattern still represents a main challenge. In fact, the number of unknowns to be retrieved is much higher than the 2D small-angle case, and a direct extension of the 2D imaging based on \emph{phase retrieval} algorithms to the 3D case \cite{raines2010three} turned out to be unreliable \cite{Barthelmes2009, wang2011non}.

To solve this \emph{dimensional deficiency}, a reduction of the parameters that describe the sample morphology is required. Up to now, wide-angle single-shot diffraction images of single particles have only been reconstructed using strongly constrained candidate geometries \cite{barke20153d, langbehn2018three, dold2020time} by \emph{fitting} the experimental data with the simulation. For example, the analysis of the diffraction data of silver nanocrystals presented in Ref. \cite{barke20153d} required the a priori identification of the main crystalline motif of the sample (tetrahedron, octahedron, icosahedron, etc.). Then, the few free parameters of the sample model, like size and orientation, were manually adjusted to fit the simulation as close as possible to the experimental data. Only as a last step, the parameters were fine-tuned automatically. Disadvantages of this procedure are (i) the results are strongly biased by the strictly constrained sample’s model, based on the prior knowledge of typical particle geometries, and (ii) already small deviations of the real structures from model shapes prevent a good agreement between measured data and forward simulations, making the search for optimal model parameters more difficult, if not impossible.

In this report we introduce a generic approach to the 3D reconstruction of faceted samples from single-shot wide-angle coherent diffraction patterns, based on a fast simulation method \cite{colombo2022scatman}, a generic model for the sample shapes and an efficient optimization strategy. We retrieve the 3D shapes of individual silver nanoparticles imaged with FEL pulses at a wavelength of \SI{5}{\nano\meter}. Even though the fitting approach does not enforce or favor any particle symmetry, regular polyhedra are retrieved in many cases, which were also reported previously \cite{barke20153d}. In addition, combinations of structural motifs and morphologies with partial or broken symmetries arise, as well as agglomerates of silver nanocrystals that were never studied so far due to the lack of suitable imaging tools. The results are presented and discussed in Sec. \ref{sec:results}, while in Sec. \ref{sec:discussion} we investigate and discuss the reliability and consistency of the 3D reconstructions and attempt to give boundaries for the spatial resolution actually achieved.

\section{Results} \label{sec:results}

\begin{figure}
\includegraphics[width=\columnwidth]{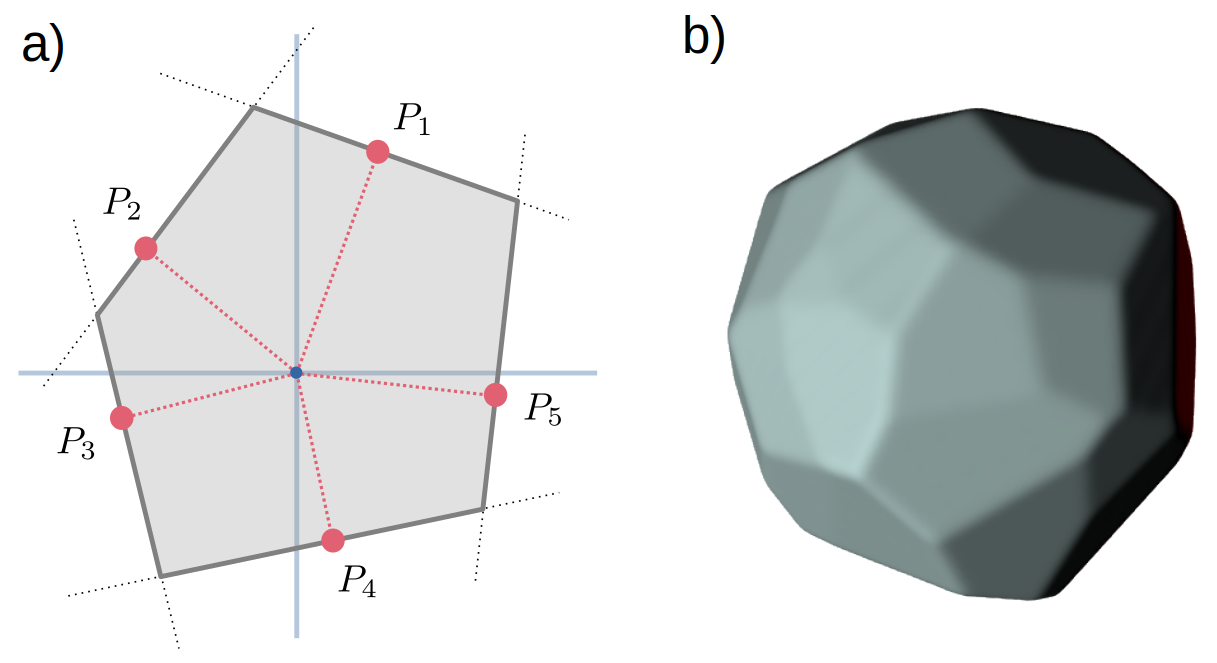}
\caption{Numerical representation of the sample's morphology. A 2D sketch of the quantities needed to define the sample's spatial extension is shown in a). A list of points $P_n$ defines $N_p$ planes, and the space enclosed by those planes defines the sample's shape. A 3D representation of a shape defined by a random positioning of the facets is shown in b).}
\label{fig:model}
\end{figure}

The experimental diffraction data \cite{dold2020time} is produced by irradiating isolated silver nanoparticles with single FEL pulses at \SI{243}{\electronvolt} (\SI{5.1}{\nano\metre}), with pulse duration between \SI{70} and \SI{120}{\femto\second}. Light is recorded up to a maximum of \SI{30}{\degree} scattering angle, equivalent to a maximum transfer momentum of \SI{0.64}{\nano\meter}$^{-1}$. 
The numerical description of the sample was chosen in a way that well accommodates the properties of silver nanoparticles \cite{barke20153d}. In fact, the shape is defined by $N_p$ planes, as sketched in Fig. \ref{fig:model}a, and is assumed to have a uniform refractive index. The imaging procedure optimizes the planes positions, which are randomly initialized as shown in Fig. \ref{fig:model}b, by forward-fitting the experimental diffraction patterns via  3D simulations \cite{colombo2022scatman}.
Further details on the experimental data and on the imaging procedure are provided in the Methods section.

\subsection{Imaging of individual nanoparticles} \label{subsec:individual}

\begin{SCfigure*}
\centering
\includegraphics[width=0.65\textwidth]{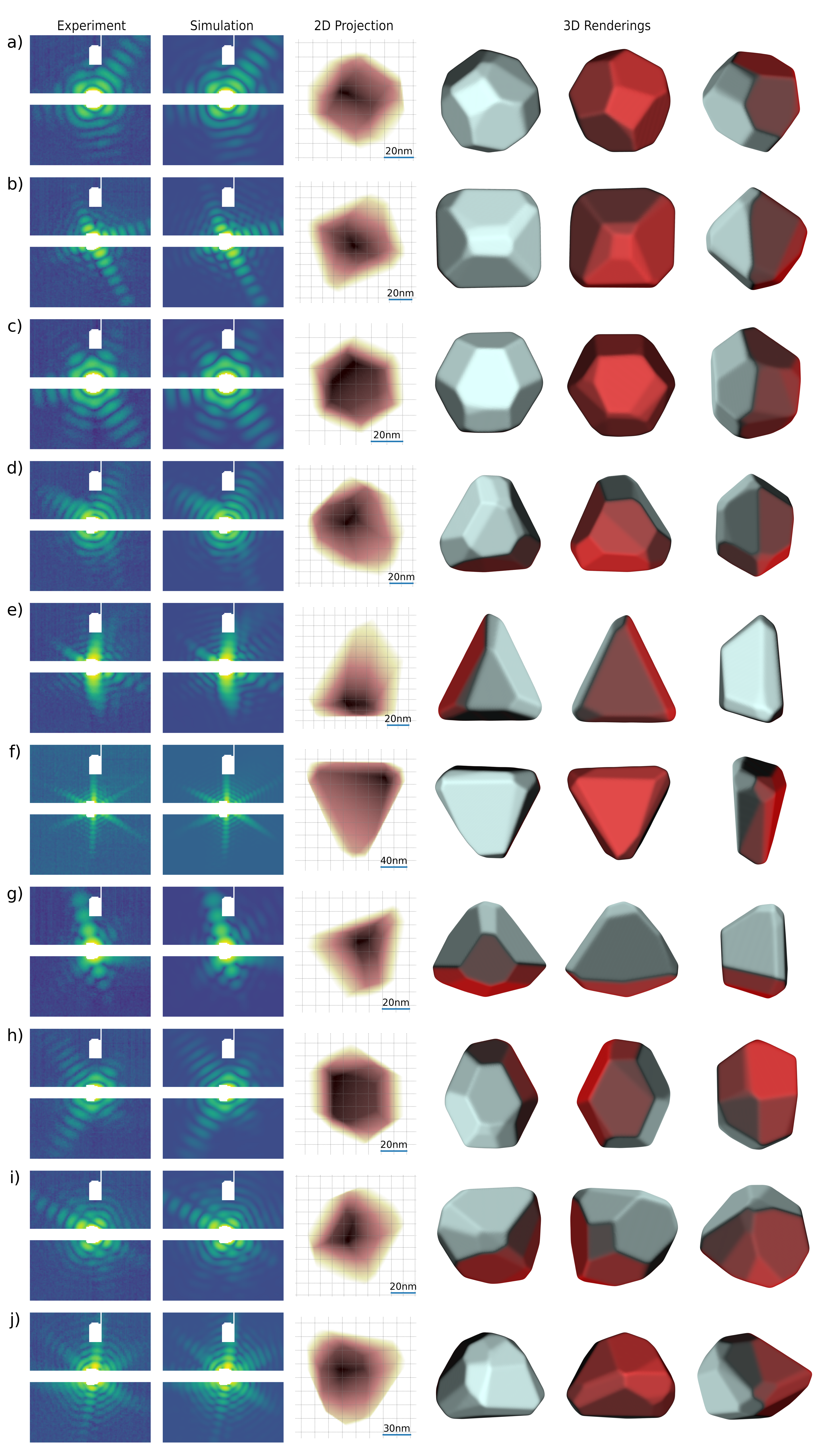}
\caption{Fitting results. Each subfigure from a) to j) is made of six columns. The first column is the experimental diffraction pattern (logarithmic color scale). The second is the retrieved diffraction pattern (logarithmic color scale). The third column is a 2D projection of the reconstruction on the plane orthogonal to the beam propagation direction. 
 The last three columns are 3D renderings of the reconstruction result. There, the sample is illuminated with white light from the \emph{beam side}, and with red light from the \emph{detector side}. For a discussion on the single subfigures, please refer to the main text.}
\label{fig:results}
\end{SCfigure*}

Fig. \ref{fig:results} shows a representative subset of the reconstructions for the presented dataset. For each reconstruction, a total of $30$ facets were initialized with random positions in space. The first column of Fig.  \ref{fig:results} depicts the experimental diffraction pattern, while the second column shows the scattering simulated from the reconstruction, both in logarithmic color scale. The experimental data is given to the fitting routine  downscaled and with saturated and missing pixels excluded. The third column shows a 2D projection of the electronic density of the reconstruction onto the detector plane (the plane orthogonal to the beam propagation direction). 
The last three columns are 3D renderings of the same reconstruction, seen from three different points of view to highlight the main features of the sample. For visualization purposes, the renderings are artificially illuminated with white light from the side that faces the incoming beam, while the opposite side, facing the detector, is lit-up in red. We note that the final number of facets in the reconstructions is, in most cases, much lower than 30, as initially given to the fitting routine. The facets that are not required to define the sample morphology are automatically placed by the fitting algorithm in a position that has no effect on the final retrieved shape.


The examples from Fig. \ref{fig:results}a to Fig. \ref{fig:results}e depict a selection out of the majority of shapes in our data set with architectures that can be directly related to known nanoparticle motifs. 
Truncated octahedra in Fig. 2a and Fig. 2b and twinned truncated tetrahedra in Fig. \ref{fig:results}c and Fig. \ref{fig:results}d with different truncation positions can be identified. Additionally, we find simple truncated tetrahedra with a strongly truncated tip, like the one shown in Fig. \ref{fig:results}e, which have not been identified in our previous experiment \cite{barke20153d}. The features in the retrieved structures underline the importance of the specific degree of truncation, which yields remarkably different architectures for the same structural motifs. The clear advantage of 3D imaging can be well appreciated by comparing the 2D projections (3rd column) in Fig. \ref{fig:results}d and \ref{fig:results}e with the 3D renderings (last columns). Due to unfortunate orientations, the symmetries of the structures can be barely identified from those 2D projections. As this is the only information that can be derived from iterative phase retrieval, the examples demonstrate that the structural properties are in many cases hardly accessible via small-angle scattering.

The second group of results, from Fig. \ref{fig:results}f to Fig. \ref{fig:results}j, are rare or unique examples from the dataset. Fig. \ref{fig:results}f is a relatively large triangular platelet which still recalls the triangular motif of the truncated tetrahedron in Fig. \ref{fig:results}e. Fig. \ref{fig:results}g also shows a retrieved nanocrystal that resembles the truncated tetrahedron. However, a small defect is present, visible on the lower side of the renderings, where a main facet is made up of two, differently oriented planes. The most pronounced effect is the deformation of the base of the truncated tetrahedron, which is no more a perfectly equilateral triangle. The retrieved architecture in Fig \ref{fig:results}h, which resembles in its main features the shape in Fig \ref{fig:results}c, actually presents two broken symmetries. Here, the twinned tetrahedron is elongated in one direction and the truncation level of the tips is different on the two sides.
Finally, Fig. \ref{fig:results}i and \ref{fig:results}j show examples for strongly asymmetric samples. The sample depicted in Fig. \ref{fig:results}i exhibits two different structural motifs. On one side, it resembles an octahedral structure like the one shown in Fig. \ref{fig:results}b, while on the opposite side a large hexagonal facet arises, similarly to the structure shown in Fig. \ref{fig:results}h. Instead, it is hard to highlight any symmetry for the reconstruction shown in Fig. \ref{fig:results}j, aside from a rough resemblance to a tetrahedral shape similar to Fig. \ref{fig:results}e and \ref{fig:results}g.

\subsection{Extension to more complex morphologies} \label{subsec:agglomerates}

\begin{figure}
\includegraphics[width=\columnwidth]{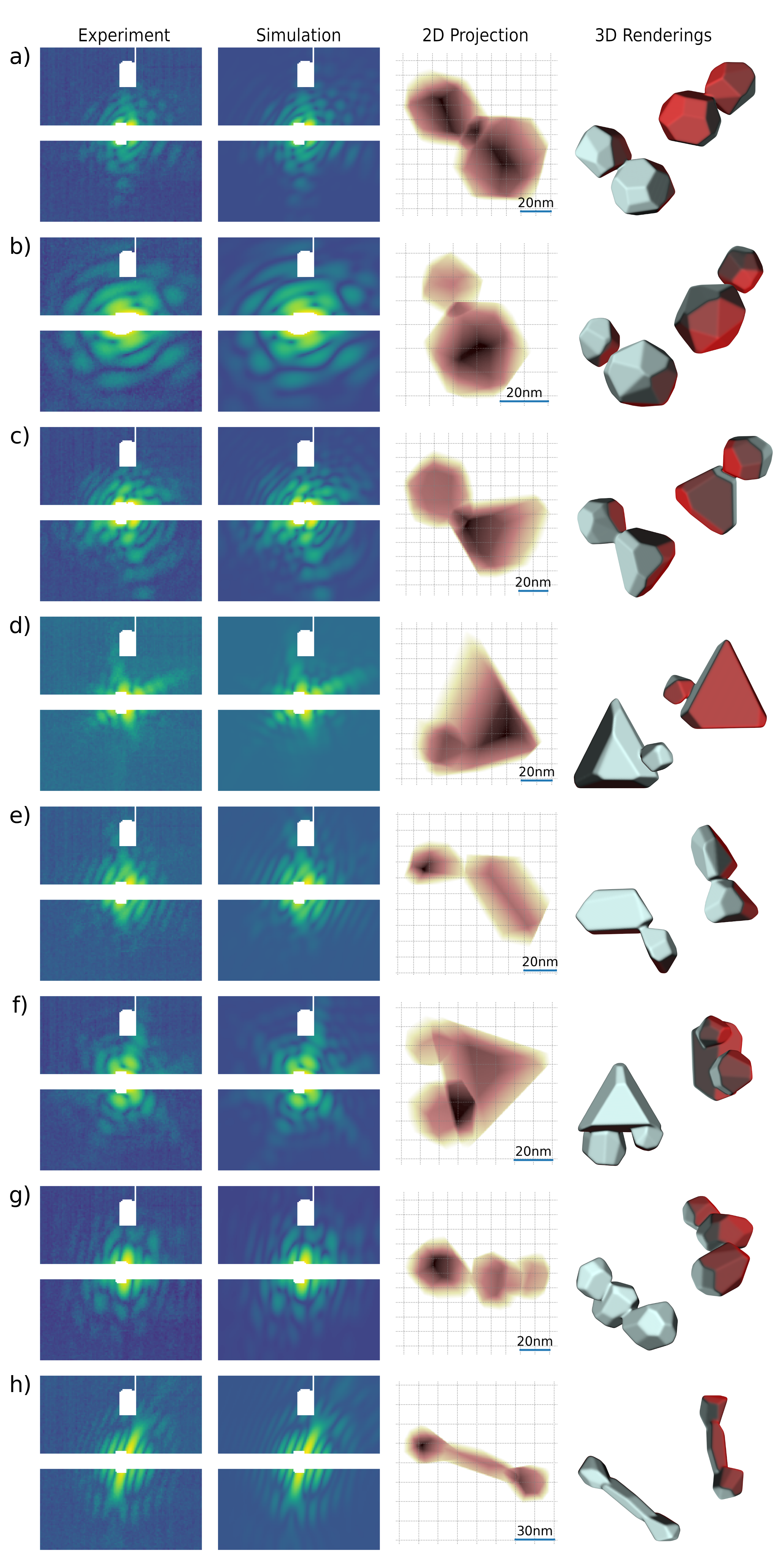}
\caption{Fitting results for silver nanocrystal agglomerates. Each subfigure from a) to h) is organized as Fig. \ref{fig:results}.}
\label{fig:results_agglomerates}
\end{figure}

When creating large nanoclusters, the formation of agglomerates \cite{huttel2017gas} is a well-known phenomenon that affects a considerable fraction of the experimental data \cite{rupp2012identification}. The presence of silver cluster agglomerates was documented in previous experiments, but the respective diffraction data were excluded from the analysis so far, due to the inability to perform a shape retrieval with the highly constrained models \cite{barke20153d}. For the analysis of patterns resulting from agglomerates, we allow the presence of multiple convex shapes properly positioned in space in a single shape model. From the numerical point of view, the optimization task becomes highly challenging, as the amount of unknowns to retrieve scales with the amount of crystals, with the addition of the three, randomly initialized, coordinates that define the relative position of each convex shape. Therefore we compose each nanoparticle of a total of 30 facets and preset the number of nanocrystals that compose an agglomerate \emph{a priori}.

A selection of the most interesting reconstructions of agglomerates is shown in Fig. \ref{fig:results_agglomerates}. The subfigures are following the scheme of Fig. \ref{fig:results}. Fig. \ref{fig:results_agglomerates}a to Fig. \ref{fig:results_agglomerates}e show agglomerates made of two subclusters, while the reconstructions from Fig. \ref{fig:results_agglomerates}f to Fig. \ref{fig:results_agglomerates}h are composed of three. We observe in general, that the structural motifs are clearly identifiable for individual clusters above \SI{50}{\nano\meter} in size. For smaller subclusters, the architecture does not present, in most cases, clear symmetries or motifs, hinting towards a possible stagnation of the optimization procedure in a local optimum, due to a number of possible reasons. First, the scattering signal is dominated by the bigger clusters in the same agglomerate. Second, the crystal itself may not present a clear structure, similar to what is observed for Fig. \ref{fig:results}i and Fig. \ref{fig:results}j. Third, the size of the single facets could be below the actual resolution limit of the diffraction data (see Sec. \ref{sec:discussion} for further discussion). 
Most of the agglomerates turn out to be composed of individual crystals whose structural motifs were also identified in reconstructions of individual clusters (Fig. \ref{fig:results}). However, some cases show peculiar features that raise the need for further discussion. We note that the bigger nanoparticle in Fig. \ref{fig:results_agglomerates}b is close to an ideal icosahedron. While  this architecture was not identified among the single nanoparticle patterns obtained in our experiment, the shape was repeatedly observed in previous experiments \cite{barke20153d}. Further, the larger structure in Fig. \ref{fig:results_agglomerates}e is peculiar, as it recalls the morphology of a trigonal platelet (see Fig. \ref{fig:results}f) with a strong truncation in one direction. The reconstruction shown in Fig. \ref{fig:results_agglomerates}h can be considered somewhat surprising. We find an agglomerate of three nanoparticles resembling the shape of a dumbbell, with two small clusters of around \SI{20}{\nano\meter} in size connected by a central column-shaped crystallite, which is the only observation of a more one-dimensional structure in our data set. 

\section{Discussion} \label{sec:discussion}

The great variety of structural motifs found in the wide-angle diffraction patterns of large silver nanoclusters, as presented in Fig. \ref{fig:results} and Fig. \ref{fig:results_agglomerates}, generally agrees well with previous observations \cite{barke20153d}. In many cases, the observed structural motifs do not correspond to the energetic ground state structures of silver clusters in this size regime, but instead give evidence of a kinematically governed growth process that gets stuck in energy minima of small seed structures in the early phase of growth \cite{barke20153d}. These meta-stable shapes together with the reconstructed architectures of the agglomerates provide a unique insight into the formation process of metal nanoparticles in the gas phase \cite{huttel2017gas}. 

We want to underline the fact that the three-dimensional information proves to be crucial for structure determination. In many cases, the retrieved 3D architectures unveil regularities that are neither intuitively expected from rather asymmetric diffraction patterns, nor easily discernible from the 2D projection of the sample density, highlighting the huge advantage gained by accessing the third spatial dimension from single diffraction images.
We further note that the ability of the approach to retrieve highly regular architectures such as the almost perfect icosahedron in Fig. \ref{fig:results_agglomerates}b or the trigonal platelet in Fig. \ref{fig:results}e is somewhat striking, as the imaging procedure does not favor any symmetry. Thus, the manifestation of these symmetries can be considered as a first strong indication of the quality of our imaging results. In addition, the three-dimensional reconstructions shown in Fig. \ref{fig:results} and Fig. \ref{fig:results_agglomerates} exhibit details that go well beyond the theoretical resolution limit of the diffraction data, raising questions about their reliability and effective resolution.

The apparent hyper-resolution is however only surprising at first glance. It arises from the presence of constraints in the definition of the shapes. In analogy, the distance of two slits can be inferred in a double slit experiment to a degree much better than the theoretical resolution limit, while the deviations of the openings from a smooth rectangular shape could not be determined better than what the resolution limit suggests. For the discussion of the accuracy of the retrieved shapes, this hyper-resolution aspect is somewhat problematic, as it renders a quantitative evaluation of the spatial resolution and the reconstruction quality very difficult. Further, in the literature, a practical demonstration of the existence of a unique solution to the 3D imaging problem from single shots is still lacking. Recently, it has been theoretically shown that a convex morphology has a unique solution \cite{engel2021modulus}, i.e., there is only one convex shape giving rise to a certain diffraction pattern. This can be seen as a strong argument for the uniqueness of the single-particle reconstructions presented in Fig. \ref{fig:results}. However, this possibly safe regime is clearly left when attempting the reconstruction of agglomerated particles.

To approach the open questions of uniqueness, reliability and resolution, in the following we provide two perspectives for discussion by (i) comparing 2D projections from our 3D reconstructions with the 2D densities obtained from well-established iterative phase retrieval using the small-angle part of the diffraction patterns and by (ii) performing a statistical analysis of the convergence properties of the 3D fitting routine.

\subsection{Comparison with iterative phase retrieval} \label{subsec:phase_ret}

\begin{figure}
\includegraphics[width=\columnwidth]{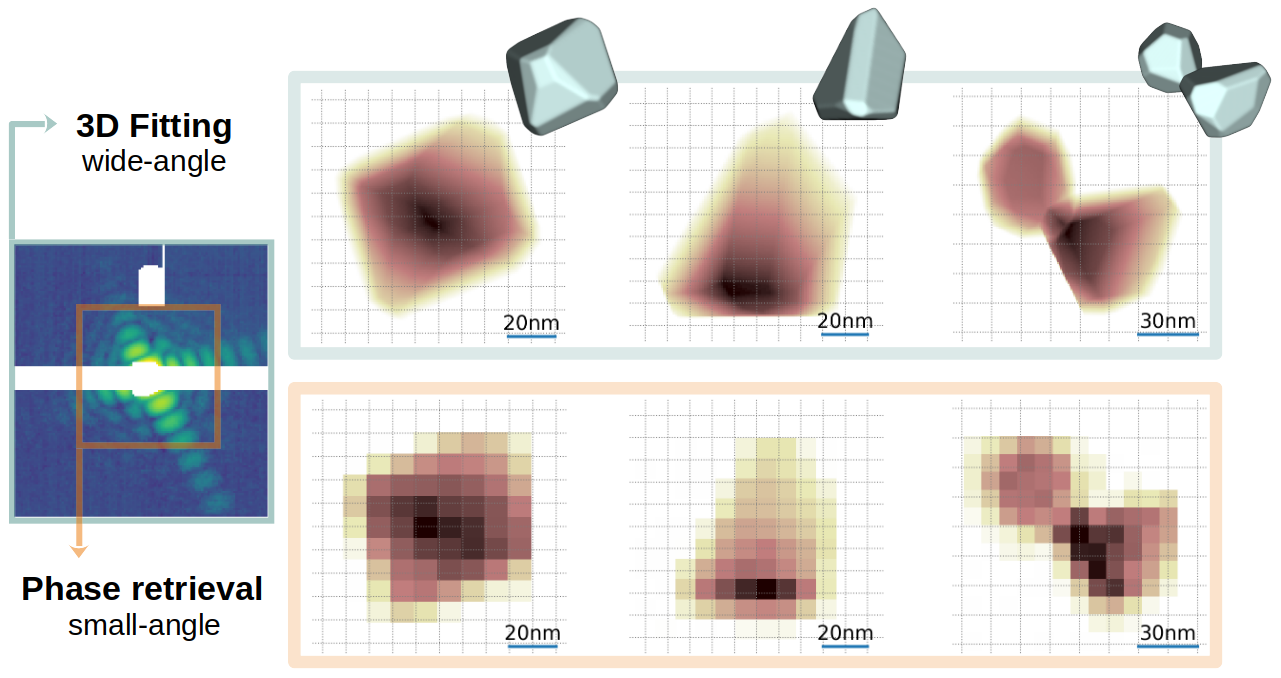}
\caption{Comparison of 3D fitting with 2D iterative phase retrieval. The comparison is performed on the experimental data presented in Fig. \ref{fig:results}b, Fig. \ref{fig:results}e and Fig. \ref{fig:results_agglomerates}c. The sample diffraction pattern on the left is overlaid by an orange box, that indicates how the diffraction data was cut, in order to select only small angle scattering signal 	on which imaging via phase retrieval algorithms was performed.}
\label{fig:phase_retrieval}
\end{figure}

In Fig. \ref{fig:phase_retrieval}, for three exemplary shapes, the 2D density projections from the corresponding three-dimensional structures are compared to 2D reconstructions derived via conventional iterative phase retrieval (IPR) \cite{marchesini2007invited} from only the small-angle part of the patterns (full and restricted small angle area up to \SI{15}{\degree} are highlighted on the left of Fig. \ref{fig:phase_retrieval}). The results of the two diﬀerent methods show an excellent agreement within the limited resolution of the IPR, thus strongly confirming the reliability of our results. The resolution of the reconstructed real-space density distribution via IPR is inherently defined by the maximum transferred momentum vector at \SI{15}{\degree}, leading to a corresponding pixel size of \SI{9}{\nano\meter}. In contrast, the resolution of the projected 3D shape appears to be much higher, even more than the theoretical limit of the full pattern of \SI{5}{\nano\meter}. As discussed above, this apparent hyper-resolution results from the constraints introduced by the shape deﬁnition via facets. Any deviation from this assumed shape, e.g. surface roughness or local features smaller than the resolution limit, will not be resolved.

\subsection{Uniqueness and Consistency} \label{subsec:uniqueness}

\begin{figure}
\includegraphics[width=\columnwidth]{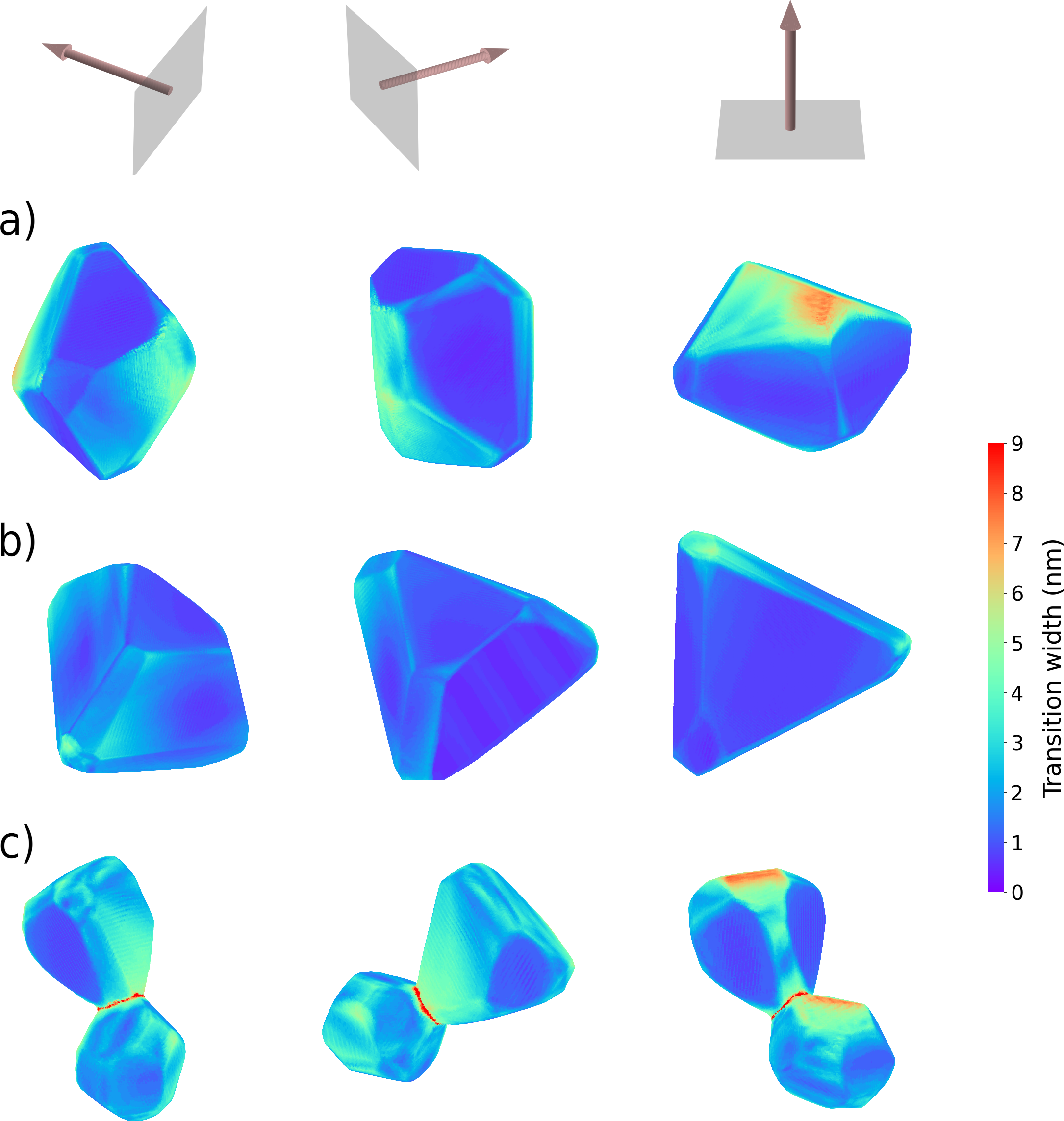}
\caption{Analysis of the reconstruction stability. In a) and b), two reconstructed samples of single nanoparticles are shown for three different orientations. These reconstructions are the average of $20$ independent fitting procedures. The orientations are indicated by the arrows on the upper row that show the beam propagation direction. For each point on the surface, a \emph{transition width} is assigned through a colormap, whose scale is indicated on the right. The given value is a measure for the stability of the reconstruction process. In c), the same analysis is presented for a nanocrystal agglomerate.}
\label{fig:stability}
\end{figure}

In order to test up to which level our results can be considered unique, it is possible to perform a statistical analysis on a set of independent reconstructions of the same diffraction data. In Fig. \ref{fig:stability}, the results for the same three shapes shown in Fig. \ref{fig:phase_retrieval} are presented.
For each structure, 20 independent imaging procedures were carried out, starting from randomly initialized facet coordinates. From the resulting 3D structures, the average density $\rho_{avg}$ was computed. Individual reconstructions are constrained to have sharp edges, i.e., they present a sudden transition between the internal density and the vacuum. However, average density distribution will not exhibit an instantaneous transition from the sample to the vacuum any more, because of the variations between the diﬀerent independent reconstructions. For each point on the average reconstruction’s edge, the transition width is computed, defined as the length of the transition from \SI{10}{\percent} to \SI{90}{\percent} of the average density profile along the surface normal. For visualization, the transition width is color-coded onto the structure surface.

Fig. \ref{fig:stability} is a strong proxy for the stability of the fitting approach in two aspects: the entity of the reconstruction uncertainty is quantified, and we can investigate how this uncertainty is distributed. First, for all three cases the transition width is much smaller than the cluster size. Second, its spatial distribution strongly varies, depending on the structures but also on the facet orientations in respect to the FEL beam. For the individual crystals (Fig. \ref{fig:stability}a and Fig. \ref{fig:stability}b), the maximum transition width is observed on those facets oriented perpendicular to the beam. In fact, the information about the placement of these facets is mostly encoded by the transfer momenta along the beam propagation direction, and thus encoded in the diffraction signal with a lower resolution.

The same considerations apply for the analysis on the nanocrystal agglomerate in Fig. \ref{fig:stability}c. Here, in  addition, an area of relatively high transition width (and thus uncertainty) is concentrated around the contact zone of the two nanocrystals. This effect can be explained by two factors. First, the computation of the \emph{transition width} where the sample's thickness approaches zero is ill-defined. Second, the contact point is in a concavity area: in such a situation, even tiny variations of the position of the facets close to the contact point have a strong effect over the size and positioning of the interface between the two nanoparticles.

Overall, the examples in Fig. \ref{fig:stability} reveal up to which detail the fitting result can be trusted, and highlight the degree of uniqueness and stability of the reconstructions. In particular, this analysis reveals uncertainties in the reconstructions that are mostly well below the \SI{5}{\nano\meter} and \SI{18}{\nano\meter} provided by the detected scattering signal on the orthogonal plane and in the depth direction, respectively.
Nevertheless, we remark that this confidence level should not be interpreted as the achieved spatial resolution of our fitting approach, because the presence of constraints introduces a strong bias that prevents a completely independent, voxel-by-voxel, resolution analysis.

\section{Summary and Outlook}

We have developed a procedure for 3D Coherent Diffractive Imaging of faceted nanoparticles employing single-shot wide-angle scattering patterns. Our analysis is based on a forward-fitting approach, where a parametrized description of the sample architecture is fitted to the experimental data. The new aspect of our method is the use of a highly generic sample parametrization, which can describe any uniform and convex sample architecture. In contrast to previous works, this feature allows to retrieve the morphology of an individual nanoparticle without constraining any symmetry, thus highly reducing the bias towards certain shapes and enormously extending the applicability of the imaging method. The results obtained for silver nanoparticles, imaged at \SI{5.1}{\nano\meter} radiation wavelength, reveal details and geometries compatible with previous experiments, along with architectures that have not been reported so far. 
The imaging method is extended to the analysis of diffraction patterns from nanocrystals agglomerates, potentially yielding a unique insight into their formation process. 
Statistical considerations on the reconstructions suggest a confidence level of the results that is well above the spatial resolution theoretically provided by the wide-angle scattering images. Furthermore, the optimization scheme does not depend on the specific parameterization and can, in principle, be extended to any other sample model.

The importance of this work, however, goes well beyond the presented reconstructions. While our demonstration focuses on faceted shapes, the optimization scheme allows for the exploration of different sample parametrizations, which will enable the study of arbitrary, even more complex structures, like complicated agglomerates or heterogenous nanoparticles as well as dynamical processes via time-resolved CDI. Although the actual resolution achieved and the limitations of our imaging approach remain to be fully defined and understood, the results presented here are a clear indication that reliable and quantitative 3D single-particle CDI from single wide-angle diffraction patterns is possible, at least under the conditions presented. The results of this work open up possibilities for wide-angle CDI that remained unexplored so far and provide an important building block for free-electron laser science towards a real 3D movie of dynamical processes at the nanoscale.


\section{Methods} \label{sec:methods}

\subsection{Experimental data} \label{sec:data}

\begin{figure}
\centering
\includegraphics[width=\columnwidth]{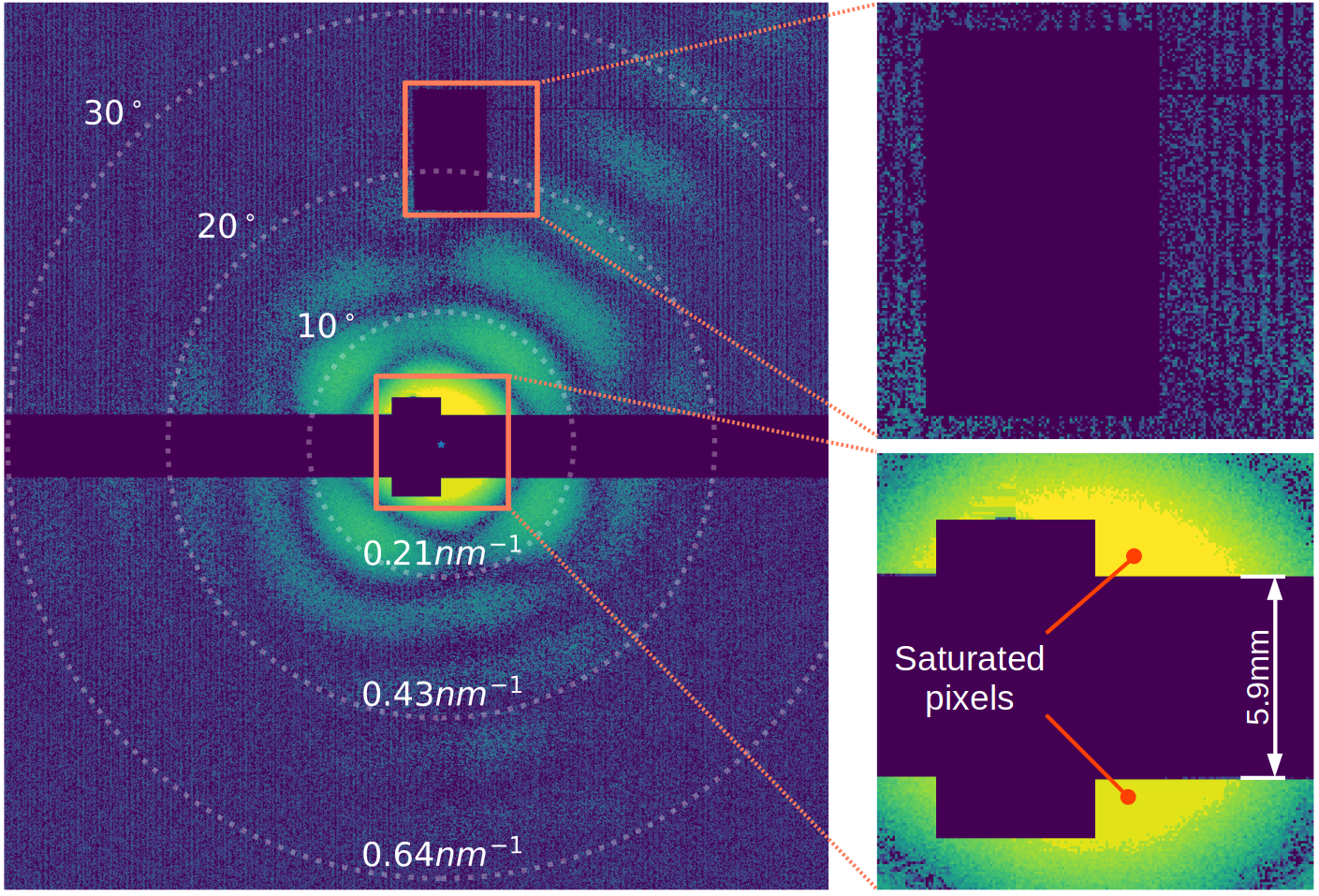}
\caption{Example of a scattering pattern. The pattern color-map is in logarithmic scale. Dotted circles indicate the scattering angles on the detector area, for $10^\circ$, for $20^\circ$ and for $30^\circ$. The corresponding values for the transferred momentum $|q|$ are also shown. The upper right inset shows peculiar features of the pnCCD detector, highlighting a rectangular defective area, excluded from the analysis, and noise. In the lower-right inset, pixels where the detector is saturated are shown in yellow. The horizontal blue stripe results from the physical separation of \SI{5.9}{\milli\metre} of the two detectors halves. Two notches are visible close to the center: their role is to reduce the risk of hitting the detector with the main X-ray beam.}
\label{fig:detector}
\end{figure}

The experimental data \cite{dold2020time} treated in this manuscript were acquired at the CAMP endstation  \cite{erk2018camp} of the Free-electron LASer in Hamburg (FLASH) \cite{ackermann2007operation}. Isolated silver nanoparticles of \SI{70}{\nano\metre} average size (equivalent to around $10^7$ silver atoms) are produced by a gas aggregation cluster source, based on the Haberland magnetron sputtersource design \cite{haberland1991new, dold2020time}, where large nano-crystals grow by coagulation of small clusters formed in gas phase from supersaturated metallic vapor.
The FEL was tuned to yield a photon energy of \SI{243}{\electronvolt}, equivalent to a wavelength of \SI{5.1}{\nano\metre}, with pulse duration between \SI{70} and \SI{120}{\femto\second}. 
The FEL beam intercepts the silver nanoparticles in front of a scattering pnCCD detector \cite{struder2010large}, composed by two halves with $1024 \times 512$ pixel resolution with a pixel size of \SI{75}{\micro\metre}$\times$\SI{75}{\micro\metre}.
The detector was positioned to record scattered light for full $2\pi$ azimuth up to $30^\circ$ scattering angle, equivalent to a maximum transfer momentum of \SI{0.64}{\nano\meter}$^{-1}$, as shown in Fig. \ref{fig:detector}.
A total of 80 diffraction patterns provided sufficient signal to be analyzed.

\subsection{The sample model} \label{sec:model}

The sample's description for this work is conveniently chosen to match the properties of silver nanoparticles \cite{barke20153d}. The shape is formed by $N_p$ planes, as depicted in Fig. \ref{fig:model}. 
Each plane is defined by the coordinates of its point closest to the origin, $P_n=\{x_n, y_n, z_n\}$. In such a way, any plane not containing the origin can be uniquely identified.
The volume enclosed by the $N_p$ planes, $V_s$, defines the sample's spatial extension. 
The optical properties of the sample are assumed to be uniform within $V_s$, i.e., the 3D spatial distribution of the refractive index $n(x,y,z)$ has a constant value $n_0$ in $V_s$, while it is set to $1$ outside.
Such a model can describe any sample that (i) does not have \emph{concavities}, (ii) is \emph{homogeneous} and (iii) has sharp boundaries. While these conditions fit well to the structural properties of silver nanocrystals \cite{barke20153d}, they are ill-suited for many other types of morphologies, such as distorted droplets, for which different parametrization basis sets may be chosen, e.g. expansion in spherical harmonics as described in Ref. \cite{colombo2022scatman}. For further considerations about the sample model, see the Supplemental Material.

\subsection{The optimization task} \label{subsec:optimization}

Our imaging approach is based on the so-called \emph{forward fitting}, where the free parameters are tuned to optimize the similarity between the simulated and the experimental pattern. The latter is downscaled through an 8x8 pixel binning, to speed up the simulation time \cite{colombo2022scatman} without having impact on the reconstruction quality, as long as the \emph{oversampling} condition is met \cite{sayre1952some}.
In our case, the similarity is encoded in the \emph{optimization target}, which has to be minimized via a suitable algorithm. Given $I^\text{exp}_{ij}$ the matrix that contains the intensity values of the experimental diffraction pattern, the optimization target $E$ is defined as:

\begin{equation}
E(\vec{p}) = \sum_{i,j} \left| I^\text{exp}_{ij} - I^\text{sim}_{ij}(\vec{p}) \right|
\label{eq:target}
\end{equation}
This equation defines the \emph{metric} used to calculate the \emph{distance} between the simulated pattern $I^\text{sim}$ and the experimental one $I^\text{exp}$. 
$I^\text{sim}$ and consequently the target $E$ are functions of the vector $\vec{p}$, which contains all the parameters that are subject to the optimization procedure. First, $\vec{p}$ contains the $3 N_p$ coordinates of the $N_p$ planes. Additionally, it encodes the complex refractive index $n$, expressed as $1 - \delta + \textbf{i}\beta$, where $\delta$ and $\beta$ are allowed to assume values in the range of $\pm$\SI{30}{\percent} of their tabulated value of $0.01$ and $0.003$ respectively \cite{henke1993x}. Finally, $\vec{p}$ includes four additional numbers: two of them are the x and y coordinates of the center of the diffraction pattern (which can vary by up to $\pm$ 2 pixel from shot to shot due to slightly different pointing directions of the FEL), one is a global offset value in the diffraction data (to account for white background noise of the pnCCD detector) and the last is a normalization factor.

For the analysis of clusters agglomerates, a number $N_s$ of faceted shapes are considered. Each  of the $N_s$ shape has its own set of facets and refractive index. Furthermore, each additional shape requires the three spatial coordinates for its relative positioning.  Thus, the total amount of free parameters $N_f$ involved in the fitting can be expressed by the following formula:
\begin{equation}
N_f = (3N_p + 2) \cdot N_s + 3\cdot(N_s-1) + 4 \, .
\label{eq:parameters}
\end{equation}
For individual nanocrystals ($N_s=1$) imaged with 30 facets ($N_p=30$), the number of parameters involved in the optimization is $N_f=96$. This number raises up to $191$ and $286$ for agglomerates composed by two and three subcrystals respectively.

Thanks to the relatively small variation from unity of the silver's refractive index at 243eV, the simulated pattern $I^\text{sim}$ as function of the parameters $\vec{p}$  can be obtained through a fast approximate simulation method, called Scatman \cite{colombo2022scatman}, capable of providing a simulated pattern in few milliseconds.
Despite the availability of a fast simulation tool, the optimization of $E(\vec{p})$ is still highly challenging, and a single reconstruction requires between 30 minutes and 2 hours to be completed on a consumer computing hardware equipped with a high-end GPU accelerator \cite{colombo2022scatman}.
The ad-hoc optimization strategy employed in this work belongs to the family of \emph{Memetic Algorithms} \cite{moscato2003gentle, colombo2017facing}. These algorithms merge stochastic and deterministic optimization techniques, aiming at combining their strengths.
In particular, a \emph{differential genetic algorithm} \cite{storn1997differential} is combined with a simplex-based \emph{Nelder-Mead algorithm} \cite{gao2012implementing}.

\subsection*{Data availability}
The analysis software has been made available online \footnote{\texttt{https://gitlab.ethz.ch/nux/numerical-physics/3d-fitting}}. The experimental data analyzed during the current study are available from the corresponding authors on reasonable request.

\vspace{2ex}

\subsection*{Acknowledgments}

We acknowledge DESY (Hamburg, Germany), a member of the Helmholtz Association HGF, for the provision of experimental facilities. The experimental campaign was carried out at FLASH via proposal F-20170541. This research was supported in part through the Maxwell computational resources operated at Deutsches Elektronen-Synchrotron DESY, Hamburg, Germany. We acknowledge the Max Planck Society for funding the development and the initial operation of the CAMP end-station within the Max Planck Advanced Study Group at CFEL and for providing this equipment for CAMP@FLASH. The installation of CAMP@FLASH was partially funded by the BMBF grants 05K10KT2, 05K13KT2, 05K16KT3 and 05K10KTB from FSP-302. 
The main financial support is kindly acknowledged from the Swiss National Science Foundation via Grant No. 200021E\_193642 and the NCCR MUST, and from DFG grant MO 719/13-1. 
Additional funding is acknowledged from the European Social Fund (ESF) and the Ministry of Education, Science and Culture of Mecklenburg-Western Pomerania (Germany) within the project NEISS under grant no ESF/14-BM-A55-0007/19.

\subsection*{Author contribution statements}

S.D., D.R., J.J, N.B, P.B., J.C., S.D., B.E., L.H., A.H., R.I., N.I., B.K., B.L, B.M., F.M, KH.MB., K.O., C.Pa., C.Pe, M.S., F.S., R.M.T., R.T., A.U., S.W., T.F., I.B., T.M. and B.I. contributed in the experiment conduction. 
A.C. contributed in developing the imaging method, performing the reconstructions and the analysis. 
A.C. and P.K. contributed in the optimization of the imaging routine.
S.D. contributed in the data selection and preparation. 
A.C, S.D, B.I and D.R. contributed in structuring and writing the manuscript, with input from all authors.

\bibliographystyle{unsrt}

\bibliography{main}

\begin{thebibliography}{10}

\bibitem{miao1999extending}
Jianwei Miao, Pambos Charalambous, Janos Kirz, and David Sayre.
\newblock Extending the methodology of x-ray crystallography to allow imaging
  of micrometre-sized non-crystalline specimens.
\newblock {\em Nature}, 400(6742):342--344, 1999.

\bibitem{chapman2010coherent}
Henry~N Chapman and Keith~A Nugent.
\newblock Coherent lensless x-ray imaging.
\newblock {\em Nature photonics}, 4(12):833--839, 2010.

\bibitem{miao2011coherent}
Jianwei Miao, Richard~L Sandberg, and Changyong Song.
\newblock Coherent x-ray diffraction imaging.
\newblock {\em IEEE Journal of selected topics in quantum electronics},
  18(1):399--410, 2011.

\bibitem{seibert2011single}
M~Marvin Seibert, Tomas Ekeberg, Filipe~RNC Maia, Martin Svenda, Jakob
  Andreasson, Olof J{\"o}nsson, Du{\v{s}}ko Odi{\'c}, Bianca Iwan, Andrea
  Rocker, Daniel Westphal, et~al.
\newblock Single mimivirus particles intercepted and imaged with an x-ray
  laser.
\newblock {\em Nature}, 470(7332):78--81, 2011.

\bibitem{ekeberg2015three}
Tomas Ekeberg, Martin Svenda, Chantal Abergel, Filipe~RNC Maia, Virginie
  Seltzer, Jean-Michel Claverie, Max Hantke, Olof J{\"o}nsson, Carl Nettelblad,
  Gijs Van Der~Schot, et~al.
\newblock Three-dimensional reconstruction of the giant mimivirus particle with
  an x-ray free-electron laser.
\newblock {\em Physical review letters}, 114(9):098102, 2015.

\bibitem{ekeberg2022observation}
Tomas Ekeberg, Dameli Assalauova, Johan Bielecki, Rebecca Boll, Benedikt~J
  Daurer, Lutz~A Eichacker, Linda~E Franken, Davide~E Galli, Luca Gelisio, Lars
  Gumprecht, et~al.
\newblock Observation of a single protein by ultrafast x-ray diffraction.
\newblock {\em bioRxiv}, 2022.

\bibitem{rupp2012identification}
Daniela Rupp, Marcus Adolph, Tais Gorkhover, Sebastian Schorb, David Wolter,
  Robert Hartmann, Nils Kimmel, Christian Reich, Torsten Feigl, Antonio~RB
  de~Castro, et~al.
\newblock Identification of twinned gas phase clusters by single-shot
  scattering with intense soft x-ray pulses.
\newblock {\em New Journal of Physics}, 14(5):055016, 2012.

\bibitem{barke20153d}
Ingo Barke, Hannes Hartmann, Daniela Rupp, Leonie Fl{\"u}ckiger, Mario Sauppe,
  Marcus Adolph, Sebastian Schorb, Christoph Bostedt, Rolf Treusch, Christian
  Peltz, et~al.
\newblock The 3d-architecture of individual free silver nanoparticles captured
  by x-ray scattering.
\newblock {\em Nature communications}, 6(1):1--7, 2015.

\bibitem{bogan2010aerosol}
Michael~J Bogan, S{\'e}bastien Boutet, Henry~N Chapman, Stefano Marchesini,
  Anton Barty, W~Henry Benner, Urs Rohner, Matthias Frank, Stefan~P Hau-Riege,
  Sasa Bajt, et~al.
\newblock Aerosol imaging with a soft x-ray free electron laser.
\newblock {\em Aerosol Science and Technology}, 44(3):i--vi, 2010.

\bibitem{langbehn2018three}
Bruno Langbehn, Katharina Sander, Yevheniy Ovcharenko, Christian Peltz, Andrew
  Clark, Marcello Coreno, Riccardo Cucini, Marcel Drabbels, Paola Finetti,
  Michele Di~Fraia, et~al.
\newblock Three-dimensional shapes of spinning helium nanodroplets.
\newblock {\em Physical review letters}, 121(25):255301, 2018.

\bibitem{bostedt2012ultrafast}
Christoph Bostedt, Ekaterina Eremina, Daniela Rupp, Marcus Adolph, Heiko
  Thomas, Matthias Hoener, Antonio~RB de~Castro, Josef Tiggesb{\"a}umker, K-H
  Meiwes-Broer, Tim Laarmann, et~al.
\newblock Ultrafast x-ray scattering of xenon nanoparticles: imaging transient
  states of matter.
\newblock {\em Physical Review Letters}, 108(9):093401, 2012.

\bibitem{fluckiger2016time}
Leonie Fl{\"u}ckiger, Daniela Rupp, Marcus Adolph, Tais Gorkhover, Maria
  Krikunova, Maria M{\"u}ller, Tim Oelze, Yevheniy Ovcharenko, Mario Sauppe,
  Sebastian Schorb, et~al.
\newblock Time-resolved x-ray imaging of a laser-induced nanoplasma and its
  neutral residuals.
\newblock {\em New Journal of Physics}, 18(4):043017, 2016.

\bibitem{Rupp2020}
Daniela Rupp, Leonie Fl{\"{u}}ckiger, Marcus Adolph, Alessandro Colombo, Tais
  Gorkhover, Marion Harmand, Maria Krikunova, Jan~Philippe M{\"{u}}ller, Tim
  Oelze, Yevheniy Ovcharenko, Maria Richter, Mario Sauppe, Sebastian Schorb,
  Rolf Treusch, David Wolter, Christoph Bostedt, and Thomas M{\"{o}}ller.
\newblock {Imaging plasma formation in isolated nanoparticles with ultrafast
  resonant scattering}.
\newblock {\em Structural Dynamics}, 7(3):34303, may 2020.

\bibitem{peltz2022few}
C~Peltz, JA~Powell, P~Rupp, A~Summers, T~Gorkhover, M~Gallei, Ina Halfpap,
  Egill Antonsson, Burkhard Langer, C~Trallero-Herrero, et~al.
\newblock Few-femtosecond resolved imaging of laser-driven nanoplasma
  expansion.
\newblock {\em New Journal of Physics}, 24(4):043024, 2022.

\bibitem{guinier1955small}
Andr{\'e} Guinier, G{\'e}rard Fournet, and Kenneth~L Yudowitch.
\newblock Small-angle scattering of x-rays.
\newblock 1955.

\bibitem{sayre1952some}
David Sayre.
\newblock Some implications of a theorem due to shannon.
\newblock {\em Acta Crystallographica}, 5(6):843--843, 1952.

\bibitem{fienup1987phase}
C~Fienup and J~Dainty.
\newblock Phase retrieval and image reconstruction for astronomy.
\newblock {\em Image recovery: theory and application}, 231:275, 1987.

\bibitem{marchesini2003x}
Stefano Marchesini, H~He, Henry~N Chapman, Stefan~P Hau-Riege, Aleksandr Noy,
  Malcolm~R Howells, Uwe Weierstall, and John~CH Spence.
\newblock X-ray image reconstruction from a diffraction pattern alone.
\newblock {\em Physical Review B}, 68(14):140101, 2003.

\bibitem{loh2012fractal}
ND~Loh, Christina~Y Hampton, Andrew~V Martin, Dmitri Starodub, Raymond~G
  Sierra, Anton Barty, Andrew Aquila, Joachim Schulz, Lukas Lomb, Jan
  Steinbrener, et~al.
\newblock Fractal morphology, imaging and mass spectrometry of single aerosol
  particles in flight.
\newblock {\em Nature}, 486(7404):513--517, 2012.

\bibitem{pedersoli2013mesoscale}
E~Pedersoli, ND~Loh, F~Capotondi, CY~Hampton, RG~Sierra, D~Starodub, C~Bostedt,
  J~Bozek, AJ~Nelson, M~Aslam, et~al.
\newblock Mesoscale morphology of airborne core--shell nanoparticle clusters:
  X-ray laser coherent diffraction imaging.
\newblock {\em Journal of Physics B: Atomic, Molecular and Optical Physics},
  46(16):164033, 2013.

\bibitem{miao2006three}
Jianwei Miao, Chien-Chun Chen, Changyong Song, Yoshinori Nishino, Yoshiki
  Kohmura, Tetsuya Ishikawa, Damien Ramunno-Johnson, Ting-Kuo Lee, and
  Subhash~H Risbud.
\newblock Three-dimensional gan- ga 2 o 3 core shell structure revealed by
  x-ray diffraction microscopy.
\newblock {\em Physical review letters}, 97(21):215503, 2006.

\bibitem{jiang2010quantitative}
Huaidong Jiang, Changyong Song, Chien-Chun Chen, Rui Xu, Kevin~S Raines,
  Benjamin~P Fahimian, Chien-Hung Lu, Ting-Kuo Lee, Akio Nakashima, Jun Urano,
  et~al.
\newblock Quantitative 3d imaging of whole, unstained cells by using x-ray
  diffraction microscopy.
\newblock {\em Proceedings of the National Academy of Sciences},
  107(25):11234--11239, 2010.

\bibitem{lundholm2018considerations}
Ida~V Lundholm, Jonas~A Sellberg, Tomas Ekeberg, Max~F Hantke, Kenta Okamoto,
  Gijs van~der Schot, Jakob Andreasson, Anton Barty, Johan Bielecki, Petr
  Bruza, et~al.
\newblock Considerations for three-dimensional image reconstruction from
  experimental data in coherent diffractive imaging.
\newblock {\em IUCrJ}, 5(5):531--541, 2018.

\bibitem{loh2010cryptotomography}
ND~Loh, Michael~J Bogan, Veit Elser, Anton Barty, S{\'e}bastien Boutet,
  Sa{\v{s}}a Bajt, Janos Hajdu, Tomas Ekeberg, Filipe~RNC Maia, Joachim Schulz,
  et~al.
\newblock Cryptotomography: reconstructing 3d fourier intensities from randomly
  oriented single-shot diffraction patterns.
\newblock {\em Physical review letters}, 104(22):225501, 2010.

\bibitem{loh2009reconstruction}
Ne-Te~Duane Loh and Veit Elser.
\newblock Reconstruction algorithm for single-particle diffraction imaging
  experiments.
\newblock {\em Physical Review E}, 80(2):026705, 2009.

\bibitem{xu2014single}
Rui Xu, Huaidong Jiang, Changyong Song, Jose~A Rodriguez, Zhifeng Huang,
  Chien-Chun Chen, Daewoong Nam, Jaehyun Park, Marcus Gallagher-Jones, Sangsoo
  Kim, et~al.
\newblock Single-shot three-dimensional structure determination of nanocrystals
  with femtosecond x-ray free-electron laser pulses.
\newblock {\em Nature communications}, 5(1):1--9, 2014.

\bibitem{raines2010three}
Kevin~S Raines, Sara Salha, Richard~L Sandberg, Huaidong Jiang, Jose~A
  Rodr{\'\i}guez, Benjamin~P Fahimian, Henry~C Kapteyn, Jincheng Du, and
  Jianwei Miao.
\newblock Three-dimensional structure determination from a single view.
\newblock {\em Nature}, 463(7278):214--217, 2010.

\bibitem{Barthelmes2009}
Franz Barthelmes.
\newblock {Definition of functionals of the geopotential and their calculation
  from spherical harmonic models : theory and formulas used by the calculation
  service of the International Centre for Global Earth Models (ICGEM)}.
\newblock Technical report, Deutsches GeoForschungsZentrum, Potsdam, sep 2009.

\bibitem{wang2011non}
Ge~Wang, Hengyong Yu, Wenxiang Cong, and Alexander Katsevich.
\newblock Non-uniqueness and instability of ‘ankylography’.
\newblock {\em Nature}, 480(7375):E2--E3, 2011.

\bibitem{dold2020time}
Simon Dold.
\newblock {\em Time-resolved imaging of laser-induced phase transitions in free
  silver nanoclusters}.
\newblock PhD thesis, Universit{\"a}t Freiburg, 2020.

\bibitem{colombo2022scatman}
Alessandro Colombo, Julian Zimmermann, Bruno Langbehn, Thomas Moller, Christian
  Peltz, Katharina Sander, Bjorn Kruse, Paul Tummler, Ingo Barke, Daniela Rupp,
  et~al.
\newblock The scatman: an approximate method for fast wide-angle scattering
  simulations.
\newblock {\em arXiv preprint arXiv:2202.03411}, 2022.

\bibitem{huttel2017gas}
Yves Huttel.
\newblock {\em Gas-phase synthesis of nanoparticles}.
\newblock John Wiley \& Sons, 2017.

\bibitem{engel2021modulus}
Konrad Engel and Bastian Laasch.
\newblock The modulus of the fourier transform on a sphere determines
  3-dimensional convex polytopes.
\newblock {\em Journal of Inverse and Ill-posed Problems}, 2021.

\bibitem{marchesini2007invited}
Stefano Marchesini.
\newblock Invited article: A unified evaluation of iterative projection
  algorithms for phase retrieval.
\newblock {\em Review of scientific instruments}, 78(1):011301, 2007.

\bibitem{erk2018camp}
Benjamin Erk, Jan~P M{\"u}ller, C{\'e}dric Bomme, Rebecca Boll, G{\"u}nter
  Brenner, Henry~N Chapman, Jonathan Correa, Stefan D{\"u}sterer, Siarhei
  Dziarzhytski, Stefan Eisebitt, et~al.
\newblock Camp@ flash: an end-station for imaging, electron-and
  ion-spectroscopy, and pump--probe experiments at the flash free-electron
  laser.
\newblock {\em Journal of synchrotron radiation}, 25(5):1529--1540, 2018.

\bibitem{ackermann2007operation}
W~al Ackermann, G~Asova, V~Ayvazyan, A~Azima, N~Baboi, J~B{\"a}hr, V~Balandin,
  B~Beutner, A~Brandt, A~Bolzmann, et~al.
\newblock Operation of a free-electron laser from the extreme ultraviolet to
  the water window.
\newblock {\em Nature photonics}, 1(6):336--342, 2007.

\bibitem{haberland1991new}
H~Haberland, M~Karrais, and M~Mall.
\newblock A new type of cluster and cluster ion source.
\newblock {\em Zeitschrift f{\"u}r Physik D Atoms, Molecules and Clusters},
  20(1):413--415, 1991.

\bibitem{struder2010large}
Lothar Str{\"u}der, Sascha Epp, Daniel Rolles, Robert Hartmann, Peter Holl,
  Gerhard Lutz, Heike Soltau, Rouven Eckart, Christian Reich, Klaus Heinzinger,
  et~al.
\newblock Large-format, high-speed, x-ray pnccds combined with electron and ion
  imaging spectrometers in a multipurpose chamber for experiments at 4th
  generation light sources.
\newblock {\em Nuclear Instruments and Methods in Physics Research Section A:
  Accelerators, Spectrometers, Detectors and Associated Equipment},
  614(3):483--496, 2010.

\bibitem{henke1993x}
Burton~L Henke, Eric~M Gullikson, and John~C Davis.
\newblock X-ray interactions: photoabsorption, scattering, transmission, and
  reflection at e= 50-30,000 ev, z= 1-92.
\newblock {\em Atomic data and nuclear data tables}, 54(2):181--342, 1993.

\bibitem{moscato2003gentle}
Pablo Moscato and Carlos Cotta.
\newblock A gentle introduction to memetic algorithms.
\newblock In {\em Handbook of metaheuristics}, pages 105--144. Springer, 2003.

\bibitem{colombo2017facing}
Alessandro Colombo, Davide~Emilio Galli, Liberato De~Caro, Francesco
  Scattarella, and Elvio Carlino.
\newblock Facing the phase problem in coherent diffractive imaging via memetic
  algorithms.
\newblock {\em Scientific reports}, 7(1):1--12, 2017.

\bibitem{storn1997differential}
Rainer Storn and Kenneth Price.
\newblock Differential evolution--a simple and efficient heuristic for global
  optimization over continuous spaces.
\newblock {\em Journal of global optimization}, 11(4):341--359, 1997.

\bibitem{gao2012implementing}
Fuchang Gao and Lixing Han.
\newblock Implementing the nelder-mead simplex algorithm with adaptive
  parameters.
\newblock {\em Computational Optimization and Applications}, 51(1):259--277,
  2012.

\bibitem{Note1}
\protect \texttt {https://gitlab.ethz.ch/nux/numerical-physics/3d-fitting}.

\end{thebibliography}

\end{document}